# Influence des lexiques d'émotions et de sentiments sur l'analyse des sentiments

## Application à des critiques de livres

**Lerch Soëlie[1], Bellot Patrice[2], Bruno Emmanuel[1] , Murisasco Elisabeth[1]**

*1. LIS UMR 7020 CNRS / AMU / UTLN, Université de Toulon*
*Campus de La Garde, 83041 Toulon Cedex 9, France*
*2. LIS UMR 7020 CNRS / AMU / UTLN, Aix Marseille Université*
*Campus de Saint Jérôme, 13 3997 Marseille Cedex 20, France*
*prenom.nom@lis-lab.fr*

RÉSUMÉ. *Les consommateurs ont l'habitude de consulter les critiques postées sur internet avant d'acheter un produit. Mais, il est difficile pour le consommateur de connaître l'opinion globale du produit vu le nombre important de ces critiques. L'analyse des sentiments permet de détecter la polarité (positive, négative ou neutre) sur une opinion exprimée et donc de classer ces critiques. Notre but est de déterminer l'influence de l'expression des émotions sur l'analyse de la polarité des critiques de livres. Nous définissons des modèles de représentation "sac de mots" de critiques qui s'appuient sur un lexique contenant des mots porteurs de sentiments (positif, négatif) et d'émotions (anticipation, tristesse, peur, colère, joie, surprise, confiance, dégoût). Ce lexique permet de mesurer les types d'émotions ressenties par les lecteurs. L'apprentissage supervisé mis en œuvre est de type forêt aléatoire (Random Forest). L'application concerne des critiques de la plateforme Amazon.*

ABSTRACT. *Consumers are used to consulting posted reviews on the Internet before buying a product. But it's difficult to know the global opinion considering the important number of those reviews. Sentiment analysis afford detecting polarity (positive, negative, neutral) in a expressed opinion and therefore classifying those reviews. Our purpose is to determine the influence of emotions on the polarity of books reviews. We define "bag-of-words" representation models of reviews which use a lexicon containing emotional (anticipation, sadness, fear, anger, joy, surprise, trust, disgust) and sentimental (positive, negative) words. This lexicon afford measuring felt emotions types by readers. The implemented supervised learning used is a Random Forest type. The application concerns Amazon platform's reviews.*

*Mots-clés : Analyse de sentiments, Analyse d'émotions (texte), Classification de polarité de sentiments*

## 1. Introduction

De nos jours, les consommateurs ont l'habitude de consulter les critiques postées sur internet avant d'acheter un produit. Cependant, le nombre d'opinions exprimées sur internet augmente considérablement. Il n'est pas évident pour le consommateur de faire le tri dans toutes ces informations pour orienter sa décision. En effet, chaque critique doit être lue en totalité pour savoir si elle est positive ou négative; cela demande donc un temps important pour connaître l'opinion globale sur un produit.

Il existe des systèmes de notation pour résumer ces critiques. Par exemple, le système de notation de la plateforme Amazon, qui met en vente des produits sur internet, propose de noter les critiques de 1 à 5 avec une échelle d'étoiles, appelée échelle de Likert (Carmines et McIver, 1981). Les critiques sont négatives si elles sont notées de 1 à 2 étoiles, positives si elles sont notées de 4 à 5 étoiles et neutres si elles sont notées 3 étoiles. Une critique neutre signifie qu'elle ne décrit aucun sentiment ou qu'elle a autant d'éléments positifs que négatifs. Ce système permet de raccourcir le temps de décision des consommateurs. Cependant, toutes les plateformes n'utilisent pas ce type de notation et le résumé de l'opinion globale du produit est difficile à définir : il faut déterminer les points positifs et négatifs d'un ensemble de critiques très important puis les comptabiliser. De plus, il n'est pas évident pour un consommateur de savoir si on peut se fier ou non à quelques critiques lues.

On peut alors avoir recours, pour classer ces critiques, à l'analyse des sentiments (Ahire *et al.*, 2017), (Liu, 2015). Elle a pour but de détecter la polarité (positive, négative ou neutre) d'un sentiment sur une opinion exprimée. Mais la présence de mots porteurs d'émotion peut être un indicateur d'informations sur ces critiques, permettant d'aider à connaître la polarité du sentiment, et mêmes les émotions véhiculées de ce texte. Cette analyse des sentiments devient donc l'analyse des émotions. La détection d'émotions est une tâche plus complexe car il s'agit de dissocier les émotions de l'auteur du texte, le ressenti du lecteur du texte et celles des éventuels personnages présents dans le texte. Par exemple, la critique suivante : "Ce livre est très triste, le héros meurt à la fin." Le terme "triste" fait référence à l'émotion "tristesse" et le terme "meurt" aux émotions "peur, colère, tristesse". Mais ces émotions négatives ne veulent pas forcément dire que la critique est négative.

Cet article propose de déterminer l'influence des émotions et des sentiments sur la polarité des critiques de livres de la plateforme Amazon. Il s'agit de pouvoir analyser cette influence sur la prédiction automatique de notes données à ces livres. Cette approche est modélisée comme un problème de classification utilisant le lexique de sentiments et d'émotions NRC (Faure, 2006). Les émotions utilisées sont : anticipation, tristesse, peur, colère, joie, surprise, confiance, dégoût et les sentiments positif, négatif.

Nos contributions consistent en :

- la création de plusieurs variantes de modèles "sac de mots" pour la création de différents modèles de représentation de données à partir d'un lexique contenant des mots porteurs de sentiments et d'émotions, pour en mesurer l'influence.

- la construction d'une méthodologie et une application de comparaison des modèles de représentation précédents s'appuyant sur la construction de modèles de prédiction de la polarité appris pour chaque modèle de représentation sur un corpus de critiques (textes isolés) grâce à un algorithme supervisé de type Random Forest. Cette méthodologie appliquée permettra de déterminer si la prédiction de la polarité des notes est meilleure avec des modèles prenant fortement en compte la présence de sentiments et d'émotions que le modèle témoin contenant tous les mots des critiques.

Cet article est organisé de la façon suivante : la section 2 présente un état de l'art sur des approches d'apprentissage statistique supervisé appliquées à l'analyse de sentiments dans les textes. La section 3 décrit les ressources utilisées dans le cadre de notre étude. La section 4 détaille notre démarche pour mesurer l'impact des mots porteurs d'émotions sur la prédiction de la polarité des critiques. La section 5 présente notre expérimentation et la sixième discute des résultats avant de conclure.

## 2. Travaux connexes

De nombreux travaux et publications ont lieu en analyse de sentiments. Par exemple, l'analyse d'une critique peut être difficile car le texte peut donner plusieurs sentiments différents sur les aspects d'un produit. Une solution possible est de créer un corpus annoté pour prédire les sentiments sur tous les aspects d'un produit (Álvarez-López et al., 2017). Une deuxième solution est l'utilisation d'un lexique de sentiments avec un score de polarité marquant l'intensité du sentiment de chaque mot (Niepert et al., 2011). (Liu et Zhang, 2012), (Bellot et al., 2015) proposent la construction de ces lexiques. Afin d'augmenter la performance des apprentissages, il est possible de déterminer l'objectivité d'une critique (neutre ou subjective), puis si elle est subjective, sa polarité (négative ou positive) (Hoffman et al., 2005). Différents modèles de représentation de données utilisant les mots comme «features» de mêmes textes donnent des résultats différents lors de l'apprentissage (Hagenau et al., 2013), (Dias et al., 2011), (Duric et Song, 2012) à partir de simples unigrammes ou bigrammes « features bas niveau » jusqu'à des mots subjectifs « features haut niveau ». Les réseaux de neurones sont utilisés dans le cas de l'analyse de sentiments pour des entités ciblées (Vo et al., 2016), ou la prédiction de polarité de messages courts tel que les tweets (Baziotis et al., 2017).

A l'instar de l'utilisation de lexiques de sentiments, des lexiques d'émotions sont utilisés pour déterminer si des phrases de critiques chinoises sont positives ou négatives (Jiao et Zhou, 2011). Une étude sur la détection de mots porteurs d'émotions dans les livres a permis de déterminer que les contes de fées font référence à une plus vaste liste d'émotions que les romans ou de calculer la fréquence de chaque émotion selon les auteurs des livres (Mohammad, 2012). L'utilisation de mots porteurs d'émotions peut donc être utile à la reconnaissance d'émotions ou de sentiments mais cela peut être insuffisant. En effet, une phrase peut aussi être construite avec des mots non porteurs de sentiments ou d'émotions; or selon son contexte, elle peut être porteuse d'émotions pour son lecteur. (Cruz-Lara *et al.*, 2010), (Cambria *et al.*, 2014) utilisent une liste de règles à appliquer sur ces phrases pour compenser ce manque d'information. Des règles peuvent être appliquées pour déterminer si deux expressions différentes expriment les mêmes émotions en utilisant le contexte des écrits (Keshtkar, 2013). (Ishizuka *et al.*, 2007) combine l'utilisation de règles et d'un lexique pour afficher les émotions d'un texte sur un avatar par des gestes et des expressions faciales .

L'analyse multimodale des sentiments exploitant du texte, des vidéos, des images et du son utilise la reconnaissance des émotions pour améliorer l'analyse des sentiments sur du texte (cambria *et al.*, 2016) ou dans le cas du "cross-domain" (Amiriparian *et al.*, 2018). Le "cross-domain" consiste à prédire la polarité des critiques sur différents types de  produits. Cela pose des difficultés dans le cas de l'analyse de sentiments sur du texte car les termes utilisés pour certains produits peuvent ne pas être utilisés pour d'autres (Bollegala *et al.*, 2013), (Bollegala *et al.*, 2016). Un thésaurus peut être nécessaire pour remplacer des mots trop spécifiques dans un domaine par un autre plus général qui traduit la même intensité du sentiment. L'analyse multimodale utilise des fichiers vocaux ou vidéos pour connaître les émotions qui peuvent être présentes pour  en déterminer la polarité. Deux mêmes émotions exprimées par la voix ou les expressions faciales sont facilement repérables tandis que des synonymes différents, dans un texte, qui expriment la même émotion sur des produits différents, sont plus difficiles à détecter.

L'analyse des émotions et l'analyse des sentiments multi-modale intéressent différents projets européens. Par exemple, le projet MixedEmotions  est un projet européen H2020 (Buitelaar, 2018)[1]. Il consiste à analyser le comportement d'un utilisateur d'un point de vue émotionnel en exploitant différentes plateformes multi-modales (texte, audio, réseaux sociaux) dans différentes langues. Plusieurs tâches ont été réalisées telles que la classification d'un texte (positif, négatif ou neutre),  d'émoticons qui  correspondent au texte et des émotions  ressenties dans un fichier texte, un fichier audio ou une vidéo. Précédemment, le projet FP7 EuroSentiment[2] (Buitelaar, 2014) a eu pour objectif la création d'une plateforme multilingue (ressources et services) pour l'analyse de sentiments. Dans un autre contexte, le projet ANR Acoformed[3] (Ochs *et al.*, 2015-2018) consiste à créer un simulateur pour entraîner les médecins à annoncer aux patients avec tact des éléments de leur maladie. Un avatar répond aux médecins en simulant les émotions que la situation transmettrait à une patiente.

[1] 1. https://mixedemotions-project.eu/, https://github.com/MixedEmotions/MixedEmotions, http://mixedemotions.insight-centre.org/
[2] 2.http://eurosentiment.readthedocs.io/en/latest/index.html
[3] 3.http://www.lpl-aix.fr/%7Eacorformed/

L'analyse des sentiments est de plus en plus utilisée pour la recommandation de produits et permet d'avoir un jugement assez fiable, notamment grâce à la présence de mots porteurs d'émotions. Plusieurs lexiques ont déjà été utilisés pour la détection d'émotions ou pour remplacer des mots pour compenser la difficulté des phrases qui semblent objectives mais qui sont porteuses d'émotions mais aucune étude n'a encore été faite pour déterminer l'impact des mots porteurs d'émotions et/ou de sentiments dans des critiques sur la polarité des sentiments. Le but de notre étude est de créer des modèles de représentation avec des mots porteurs de sentiments et/ou d'émotions en se basant sur un lexique pour déterminer si l'impact des sentiments et/ou des émotions sur les critiques améliorent la prédiction des critiques par un algorithme d'apprentissage de type Random Forest.

## 3. Ressources

Afin de déterminer si la présence des émotions et des sentiments permettent d'influencer la prédiction de notes de critiques, nous proposons d'utiliser des lexiques d'émotions et de sentiments pré-construits. Quelques lexiques existent pour l'anglais tels que le lexique NRC (Faure, 2006) et pour le français une liste de mots par émotions Voie-de-l'écoute (Sanchez, 2014) . Plusieurs corpus existent pour l'entraînement existent tels que le corpus de SemEval (Strapparava et Mihalcea, 2007), (Strapparava et Mihalcea, 2010) ou le corpus de critiques de la plateforme d'Amazon (Bogers *et al.,* 2016).

### 3.1. Lexiques d'émotions et de sentiments

Pour créer les modèles de représentation des critiques, une étude sur les lexiques d'émotions et de sentiments est nécessaire. Nous avons choisi le lexique NRC[4] car il contient un important nombre de mots porteurs de sentiments et d'émotions, il a déjà également été utilisé dans une étude de fréquence de mots porteurs d'émotions dans du texte (Mohammad, 2012). Celui-ci contient un terme par ligne suivi d'une émotion ou d'un sentiment parmi la liste suivante "anticipation, colère, tristesse, joie, peur, confiance, dégoût, positive, négative" et d'un 0 ou d'un 1. S'il y a 1, cela veut dire que le mot est porteur de l'émotion sinon cela veut dire que le mot ne l'est pas. Ce lexique contient 13668 mots associés à un sentiment et/ou une émotion. Il est également possible qu'un même mot soit associé à plusieurs émotions. Lors de la définition des modèles, ce lexique nous servira de support pour enrichir les descripteurs (*features*) utilisés par les classifieurs.

---

[4].https://github.com/sebastianruder/emotion_proposition_store/tree/master/NRC-Emotion-Lexicon-v0.92

Par exemple, le mot "*abandon*" dans le lexique NRC est associé aux émotions de peur et de tristesse et au sentiment négatif mais n'est pas associé aux autres émotions de colère, d'anticipation, de dégoût, de joie, de surprise et de confiance ni au sentiment positif.

### 3.2. Corpus de critiques de livres d'Amazon

Le corpus de critiques de livres de la plateforme Amazon (Bogers *et al.,* 2016) contient le nombre total de critiques sous forme de fichiers XML pour les 2,8 millions de livres disponibles. Il contient des informations telles que le code ISBN du livre, son titre, ses auteurs… et les critiques du livre ainsi que la note associée à chaque critique. La taille du corpus est de 29Go. Il s'agit d'un corpus qui a été utilisé dans les évaluations internationales en recherche d'information CLEF Social Book Search[5]. Il contient plus d'une dizaine de millions de critiques telle que celle-ci, avec la note associée :

<content>*Interesting Grisham tale of a lawyer that takes millions of dollars from his firm after faking his own death. Grisham usually is able to hook his readers early and ,in this case, doesn't play his hand to soon. The usually reliable Frank Mueller makes this story even an even better bet on Audiobook.*</content>

<rating>4</rating>

### 4. Méthodologie

Dans cette section, nous utilisons une représentation "sac de mots" pour déterminer si la présence des émotions permettent de mieux prédire la polarité des critiques. Les algorithmes de classification utilisent un modèle contenant tous les mots des critiques. Nous proposons ici plusieurs modèles de représentation des critiques qui permettent de déterminer si les modèles utilisant les mots porteurs de sentiments et d'émotions sont meilleurs que les mots porteurs seulement de sentiments ou porteurs seulement d'émotions. Les émotions que nous considérons sont "anticipation, joie, colère, tristesse, dégoût, confiance, peur, surprise". Les sentiments sont représentés par leur polarité : positif, négatif. Nous utilisons un lexique pour représenter un mot par une émotion, un corpus afin d'entraîner un classifieur de type Random Forest. Cet algorithme de classification permet d'ordonner les features plus ou moins discriminantes (les termes les plus fréquents). Nous expliquons comment évaluer l'algorithme d'apprentissage avec les modèles cités précédemment.

---

[5] 5. http://social-book-search.humanities.uva.nl/

### 4.1. Modèles de représentation

Nous proposons et expérimentons un certain nombre de modèles de représentation des données, partant d'un modèle de base qui contient tous les mots des critiques (Modèle M ci-dessous) jusqu'à des modèles plus complexes. Toutes les ponctuations, majuscules et caractères spéciaux sont supprimés. Pour illustrer ces modèles, nous partons de l'exemple ci-dessous (les mots soulignés dans le texte sont présents dans le lexique de sentiments et d'émotions) :

*"Interesting Grisham tale of a lawyer that takes millions of dollars from his firm after faking his own death. Grisham usually is able to hook his readers early and, in this case, doesn't play his hand to soon. The usually reliable Frank Mueller makes this story even an even better bet on Audiobook."*

– **Modèle Mots (M) :** contient tous les mots des critiques : "*interesting grisham tale of a lawyer that takes millions of dollars from his firm after faking his own death grisham usually is able to hook his readers early and in this case doesn't play his hand to soon the usually reliable frank mueller makes this story even an even better bet on audiobook*".

– **Modèle Emotions Sentiments (ES) :** ne sont conservés que les mots porteurs d'émotions ou de sentiments présents dans le lexique de sentiments et d'émotions. L'exemple devient alors : "*interesting death hook play reliable better*".

– **Modèle Sentiments (S):** ne sont conservés que les mots porteurs de sentiments présents dans le lexique de sentiments et d'émotions. Cela ne veut pas dire que tous les mots porteurs d'émotions sont supprimés, seulement ceux qui ne font référence à aucun sentiment. L'exemple devient alors : "*interesting death hook play reliable better*". Dans cet exemple précis, celui-ci devient exactement le même que le modèle ES mais cela n'est pas une généralité. Cela est vrai dans le lexique mais en réalité, des termes qui font référence à une émotion semblent obligatoirement faire référence au minimum à un sentiment. On peut se poser la question de présence d'émotion neutre tel que la "surprise" qui peut être "positive" ou "négative".

– **Modèle Emotions (E) :** ne sont conservés que les mots porteurs d'émotions présents dans le lexique de sentiments et d'émotions. Cela ne veut pas dire que tous les mots porteurs de sentiments sont supprimés, seulement ceux qui ne font référence à aucun sentiment. L'exemple devient alors : "*death hook reliable better*".

– **Modèle Emotions Sentiments + mots Génériques (ES+G) :** les mots qui ne sont pas présents dans le lexique de sentiments et d'émotions sont remplacés par le mot générique "*non_emotion*". L'exemple devient "*interesting non_emotion non_emotion[...] death non_emotion [...] hook non_emotion [...] play non_emotion [...] reliable non_emotion [...] better non_emotion [...]*".

– **Modèle Sentiments + mots Génériques (S+G) :** les mots qui ne sont pas présents dans le lexique de sentiments et qui sont porteurs d'émotions mais pas de sentiments sont remplacés par le mot générique "*non_emotion*". L'exemple devient "interesting *non_emotion* non_emotion[...] death *non_emotion* [...] hook *non_emotion* [...] play *non_emotion* [...] reliable *non_emotion* [...] better *non_emotion* [...]".

– **Modèle Emotions + mots Génériques E+G :** les mots qui ne sont pas présents dans le lexique de sentiments et qui sont porteurs de sentiments mais pas d'émotions sont remplacés par le mot générique "*non_emotion*". L'exemple devient "*non_emotion* non_emotion[...] death *non_emotion* [...] hook *non_emotion* [...] reliable *non_emotion* [...] better *non_emotion* [...]".

– **Modèle Catégorie d'Emotions et de Sentiments + autre Mots ($C_{ES}$ + M) :** les mots porteurs d'émotions ou de sentiments présents dans le lexique de sentiments et d'émotions sont remplacés par les noms des catégories des émotions ou des sentiments qui leur correspondent. L'exemple devient "*positive grisham tale of a lawyer that takes millions of dollars from his firm after faking his own anger sadness fear negative grisham usually is able to positive joy his readers early and in this case doesn't positive his hand to soon the usually trust positive frank mueller makes this story even an even positive joy bet on audiobook*".

– **Modèle Catégorie de sentiments + autres Mots ($C_S$ + M):** les mots porteurs de sentiments présents dans le lexique de sentiments et d'émotions sont remplacés par les noms des catégories des sentiments qui leur correspondent. L'exemple devient "*positive grisham tale of a lawyer that takes millions of dollars from his firm after faking his negative is able to positive his readers early and in this case doesn't positive his hand to soon the usually positive frank mueller makes this story even an even positive bet on audiobook*".

– **Modèle Catégorie d'Emotions + autres Mots ($C_E$ + M):** les mots porteurs d'émotions présents dans le lexique de sentiments et d'émotions sont remplacés par les catégories des émotions qui leur correspondent. L'exemple devient"*Interesting grisham tale of a lawyer that takes millions of dollars from his firm after faking his own anger sadness fear grisham usually is able to joy his readers early and in this case doesn't play his hand to soon the usually trust frank mueller makes this story even an even joy bet on audiobook*".

– **Modèle M sans Emotions et sans Sentiments (M — ES) :** les mots porteurs d'émotions et de sentiments sont supprimés. L'exemple devient "*grisham tale of a lawyer that takes millions of dollars from his firm after faking his own grisham usually is able to his readers early and ,in this case doesn't his hand to soon the usually frank mueller makes this story even an even bet on audiobook.*"

### 4.2. Evaluation des apprentissages des modèles

Ces modèles sont évalués par trois types de mesure *MicroMoyennePrécision*, *MicroMoyenneRappel*, *MicroMoyenneF-mesure*. Cette démarche permet de mesurer l'impact des mots porteurs d'émotions sur la prédiction des notes des critiques. Il est difficile de prédire la vraie note car différentes personnes qui ont le même avis ne notent pas en utilisant la même échelle. Il est plus pertinent de mesurer la polarité. C'est pour cela que nous utilisons seulement 3 classes : "1,2" (notes négatives), "3" (note neutre), "4,5" (notes positives) au lieu de 5 classes : "1", "2", "3", "4", "5" .

A partir du corpus d'Amazon, les critiques sont extraites des fichiers XML et un modèle "sac de mots" est construit, où chaque critique est représentée par une suite de mots (Modèle M). Grâce au lexique d'émotions et de sentiments, les critiques sont représentées sous la forme de différents modèles de représentation. Ces "sacs de mots" servent de données d'entraînement au classifier qui construit des représentations vectorielles. Les mots présents cité zéro ou une fois dans une critique, sont supprimés pour réduire la dimension des modèles et limiter le risque de sur-apprentissage sur des informations trop rares. L'entraînement des données est utilisé avec un classifier de type forêt aléatoire (Random Forest) avec les paramètres par défaut de l'algorithme de Scikit Learn[6] (profondeur, nombre de séparations à chaque noeud...) Les arbres se construisent avec un calcul de gain en entropie. La séparation d'un noeud en deux noeuds se fait selon un seuil de fréquence des mots. L'évaluation est faite avec la validation croisée en 10 plis.

## 5. Expérimentation

Cette section présente les expérimentations effectuées sur le jeu de données Amazon, l'objectif étant de prédire automatiquement les notes données par les lecteurs en fonction des critiques qu'ils ont écrites. Les programmes qui ont permis d'obtenir ces résultats ont été lancés sur un serveur de calcul Dell PowerEdge R730 avec 64Go de RAM et deux processeurs Intel(R) Xeon(R) E5-2630 v3 à 2.40GHz avec chacun 8 coeurs exécutant 2 threads. Appliquant la validation croisée sur 10 plis sur un classifier de type Random Forest, une représentation de type "sac de mots" pour chaque modèle est choisie après avoir été transformée en vecteurs de fréquence pour chaque mot.

Le nombre de critiques  dans chaque classe pour l'entraînement n'est pas toujours le même. D'après (Martin, 2017),  différents types d'échantillonnage font varier les résultats des classifieurs. Les trois répartitions sont le sous-échantillonnage des classes majoritaires, le sur-échantillonnage des classes minoritaires et la répartition inégale des classes. Si aucune critique n'est supprimée ou ajoutée, la répartition des classes est par défaut inégale. C'est à dire que le nombre de critiques dans chaque classe n'est pas identique. Pour 848399 critiques, il y en a 13% dans la classe "1,2", 9% dans la classe "3" et 78% dans la classe "4,5". Pour répartir équitablement les données, il existe deux méthodes:

– Les classes majoritaires peuvent être sous-échantillonnées : le nombre de critiques dans chaque classe est compté et tant que le nombre de critiques des classes majoritaires n'est pas égal au nombre de critiques de la classe minoritaire, les critiques en surplus des classes majoritaires sont supprimées.

– Les classes minoritaires peuvent être sur-échantillonnées : le nombre de critiques dans chaque classe est compté et tant que le nombre de critiques des classes minoritaires n'est pas égal au nombre de critiques de la classe majoritaire, des critiques des classes minoritaires sont copiées.

---

[6] https://scikit-learn.org/stable/modules/generated/sklearn.ensemble.RandomForestClassifier.html

(Martin, 2017) montre que les résultats changent en fonction de ces répartitions. Notre étude a été faite avec ces trois répartitions de données. Avec le sous-échantillonnage des données, les résultats sont proches de l'aléatoire. Avec la répartition naturelle inégale, si nous comparons les résultats avec ceux que nous avons si la classe majoritaire est prédite pour toutes les données de test, alors nous avons des résultats quasiment identiques. C'est pour cela que les résultats ne seront présentés ici que dans le cas du sur-échantillonnage (Tableau 1).

*Tableau 1. Résultats en fonction des différents modèles de représentation et du nombre des critiques sur 983040 critiques*

| Métriques | MicroMoyenne Précision | MicroMoyenneRappel | MicroMoyenneF-mesure |
|---|---|---|---|
| ModèleM | **0,93** | **0,92** | **0.92** |
| ModèleSE | 0,842 | 0,831 | 0.83 |
| ModèleS | 0,807 | 0,799 | 0.792 |
| ModèleE | 0,78 | 0,77 | 0.77 |
| ModèleSE+G | 0,88 | 0,871 | 0.87 |
| ModèleS+G | 0,86 | 0,85 | 0.85 |
| ModèleE+G | 0,841 | 0,839 | 0.832 |
| ModèleCSE+M | **0,93** | **0,92** | **0.92** |
| ModèleCS+M | **0,93** | **0,92** | **0.92** |
| ModèleCE+M | **0,93** | **0,92** | **0.92** |
| Modèle — SE | 0,921 | **0,92** | **0.92** |

## 5. Discussion

D'après les résultats, les modèles prenant en compte tous les mots porteurs de sentiments et d'émotions sont plus performants que ceux qui prennent en compte seulement les mots porteurs de sentiments ou seulement les mots porteurs d'émotions. Il n'y a qu'une exception à la règle : ceux qui gardent les mots qui ne sont pas porteurs de sentiments ni d'émotions et qui remplacent les autres mots par leur nom de catégorie de sentiments et d'émotions (CSE+M, CS+M, CE+M). Cela est lié par le fait de la présence de mots qui ne sont pas dans le lexique de sentiments et d'émotions. La différence entre les résultats de ces modèles particuliers est très faible voire nulle si le nombre de critiques est important. Ces résultats sont d'ailleurs les mêmes que le modèle qui contient tous les mots et le modèle qui ne contient aucun mot porteur de sentiments et d'émotions. On peut se poser la question de la présence de mots qui n'ont pas de significations sentimentales ou émotionnelles particulières mais qui sont plus souvent utilisées dans les critiques positives que les critiques négatives ou neutres et vice-versa ou également si certaines expressions sont porteuses d'émotions mais pas les mots eux même comme le montre (Cruz-Lara *et al.*, 2010). En comparant les modèles (SE+G, S+G, E+G) et les modèles (SE, S, E), il est évident que le nombre de termes des critiques jouent également un rôle dans la prédiction des sentiments et des émotions car la seule différence entre les deux est que dans le cas des modèles (SE+G, S+G, E+G), tous les mots qui ne sont pas porteurs de sentiments et d'émotions sont remplacés par un mot générique qui n'a aucun sens pour l'algorithme d'apprentissage.

## 7. Conclusion et futurs travaux

Cette étude est basée sur le problème de la prédiction des notes en fonction de l'influence des émotions. Onze modèles de représentation des données en différents "sacs de mots" sont proposés. Les expérimentations permettent de démontrer que les mots porteurs de sentiments et d'émotions permettent de mieux apprendre que les mots porteurs de sentiments ou d'émotions suite à un sur-échantillonnage de la classe minoritaire. Une telle approche nécessite donc une connaissance a priori des répartitions des différentes classes de sentiment du corpus pour être réalisable. Les modèles d'apprentissage pourront être enrichis par la représentation vectorielle continue des mots.

Dans une future étude, nous souhaitons prédire les émotions associées à ces critiques au lieu de prédire la polarité de ces critiques. Pour cela, il faudra construire un corpus annoté émotion. Cette tâche est très complexe car trois types d'émotions peuvent êtres rencontrés dans les critiques en plus des catégories d'émotions déjà présentées : celles ressenties par l'auteur de la critique, celles que l'auteur de la critique veut faire ressentir, mais aussi, dans le cas d'un livre de fiction par exemple, celles des personnages de l'histoire. Nous n'avons pas de critiques annotées de cette manière. Nous commençons donc par produire un guide et un langage d'annotation pour les annoter. Mais pour cela, il faudra donc plusieurs annotateurs sur les mêmes critiques pour pouvoir les valider. Une fois le corpus construit, un classifieur pourra être entraîné pour arriver à deviner les émotions ressenties par le lecteur du livre ou également quelles émotions font qu'un livre est aimé ou non par un lecteur. Nous pourrons également utiliser d'autres modèles de représentation des données tels que le plongement de mots et également d'autres type d'apprentissages tel que les réseaux de neurones